%% file: bare_jrnl_new_sample4.tex
\documentclass[lettersize,journal]{IEEEtran}
\usepackage{amsmath,amsfonts}
\usepackage{algorithmic}
\usepackage{algorithm}
\usepackage{array}
\usepackage[caption=false,font=normalsize,labelfont=sf,textfont=sf]{subfig}
\usepackage{textcomp}
\usepackage{stfloats}
\usepackage{url}
\usepackage{verbatim}
\usepackage{graphicx}
\usepackage{cite}
\usepackage{booktabs}
\usepackage{multirow}
\usepackage{enumitem}
\usepackage{amsmath}
\usepackage{amsfonts}
\usepackage{subfig}
\usepackage{color}

\begin{document}

\title{A deep local attention network for pre-operative lymph node metastasis prediction in pancreatic cancer via multiphase CT imaging}

\author{Zhilin Zheng, Xu Fang, Jiawen Yao, Mengmeng Zhu, Le Lu ~\IEEEmembership{Fellow,~IEEE,} Lingyun Huang, Jing Xiao, Yu Shi, Hong Lu, Jianping Lu, Ling Zhang, Chengwei Shao*, Yun Bian*
\thanks{Z. Zheng, L. Huang, J. Xiao are with Ping An Technology (Shanghai \& Shenzhen), People’s Republic of China, (e-mail: zhilin.zheng95@gmail.com).}
\thanks{X. Fang, M. Zhu and J. Lu are with Changhai Hospital, Shanghai, People’s Republic of China. J. Yao, L. Lu and L. Zhang were with PAII Inc., Bethesda, MD 20817, USA. X. Fang and J. Yao contributed equally.}
\thanks{Y. Shi is with Department of Radiology, Shengjing Hospital of China Medical University, Shenyang, China.}
\thanks{H. Lu is with Department of Radiology, Tianjin Medical University Cancer Institute and Hospital, National Clinical Research Center of Cancer, Key Laboratory of Cancer Prevention and Therapy, Tianjin, China.}
\thanks{*s indicate joint corresponding authors. Y. Bian and C. Shao are with Changhai Hospital, Shanghai, People’s Republic of China, (email: bianyun2012@foxmail.com, cwshao@sina.com).}

}



\maketitle

\begin{abstract}
Lymph node (LN) metastasis status is one of the most critical prognostic and cancer staging factors for patients with resectable pancreatic ductal adenocarcinoma (PDAC), or in general, for any types of solid malignant tumors. Preoperative prediction of LN metastasis from non-invasive CT imaging is highly desired, as it might be straightforwardly used to guide the following neoadjuvant treatment decision and surgical planning. Most studies only capture the tumor characteristics in CT imaging to implicitly infer LN metastasis and very few work  exploit direct LN's CT imaging information. LN staging is usually confirmed from pathological images acquired after invasive procedures of biopsy or surgery. To the best of our knowledge, this is the first work to propose a fully-automated LN segmentation and identification network to directly facilitate the LN metastasis status prediction task. Nevertheless LN segmentation/detection is very challenging since LN can be easily confused with other hard negative anatomic structures (e.g., vessels) from radiological images. 1) We explore the anatomical spatial context priors of pancreatic LN locations by generating a guiding attention map from related organs and vessels to assist segmentation and infer LN status. As such, LN segmentation is impelled to focus on regions that are anatomically adjacent or plausible with respect to the specific organs and vessels (thus hard negative samples with certain distance ranges can be ignored). 2) The metastasized LN identification network is trained to classify the segmented LN instances into positives or negatives by reusing the segmentation network as a pre-trained backbone and padding a new classification head. 3) More importantly, we develop a LN metastasis status prediction network that combines the patient-wise aggregation results of LN segmentation/identification and deep imaging features extracted from the tumor region. 4) Extensive quantitative nested five-fold cross-validation is conducted on a discovery dataset of 749 patients with PDAC. External multi-center clinical evaluation is further performed on two other hospitals of 191 patients in total. Our final multi-staged LN status prediction network statistically significantly outperforms the strong baseline of nnUNet and several other compared methods, including CT-reported LN status, radiomics, and deep learning models.
\end{abstract}


\section{Introduction}\label{sec:introduction}
\IEEEPARstart{P}{ancreatic} cancer is the third leading cause of overall cancer death in the United States\cite{siegel2021cancer}, of which approximately 95$\%$ is pancreatic ductal adenocarcinoma (PDAC)\cite{grossberg2020multidisciplinary}. With the poorest prognosis (i.e., 5-year overall survival (OS) of 10$\%$ approximately), surgical resection is the most effective way to achieve long-term survival for  patients with PDAC\cite{grossberg2020multidisciplinary}. However, not all patients can benefit from the margin-negative (R0) resection and comprehensive treatment protocol is usually established for pancreatic cancer. The patients' treatment selections can be determined by whether their peripancreatic lymph nodes (LNs) have metastasized with the options of adjuvant radiotherapy (RT) or chemotherapy. 
It was found that neoadjuvant therapy before surgery was associated with improved survival and time to recurrence in patients with LN metastasis, since neoadjuvant therapy can not only treat  lymphovascular invasion but also benefit tumor downstaging \cite{roland2015neoadjuvant,kanda2011pattern}. 
The accurate preoperative detection of LN metastasis becomes vital and would aid in treatment planning and management. 

Contrast-enhanced CT is used as the typical imaging protocol for identifying the presence of peripancreatic metastatic disease to LNs, but it is a very challenging task for radiologists to determine whether a patient has LN metastasis by using only CT scans. To this end, poor diagnostic accuracy of CT with a pooled sensitivity of 25$\%$ and positive predictive value (PPV) of 28$\%$ was reported in a meta-analysis \cite{tseng2014diagnostic} on assessing extra-regional LN metastasis in pancreatic and peri-ampullary cancer. 
Recently, several radiomics based approaches have been proposed to tackle the LN metastasis differentiation problem of various cancer types ~\cite{ji2019biliary,wang2020ct,li2020contrast,bian2019relationship,yang2020integrating,gao2020radiomics,meng20202d}. However, these methods require hand-crafted feature design which can bring concerns of reproducibility and human bias is often introduced due to manual selection of 2D tumor slice with limited representation power. Although there are some deep learning work that report promising performance on predicting LN metastasis status in gastric cancer~\cite{jin2021deep, dong2020deep}, those models assume that the risk of metastases is fundamentally driven by the primary tumor. They rely on LN CT report information for the integration model without using any LNs detection or segmentation. The model that takes both tumor morphology and lymphatic anatomy into account could be of more clinically usefulness on addressing these aforementioned issues, similarly as in the diagnostic processes performed by radiologist readers. PET/CT is another imaging modality worth exploring. PET/CT based approaches \cite{kim2019predictive,asagi2013utility,dahmarde202018f} generally use maximum standardized uptake value ($SUV_{max}$) of manually-drawn LN RoIs as the prediction element, but it comes with challenges of numerous false positives from inflammatory LNs and false negatives from small-sized metastatic LNs \cite{tseng2021role, jung2017value}. Also, it is relatively not as common as CT, which is less affordable, available and accessible, hence we opt for CT for our research purpose.

\begin{figure*}[htb]
\centerline{\includegraphics[width=0.9\linewidth]{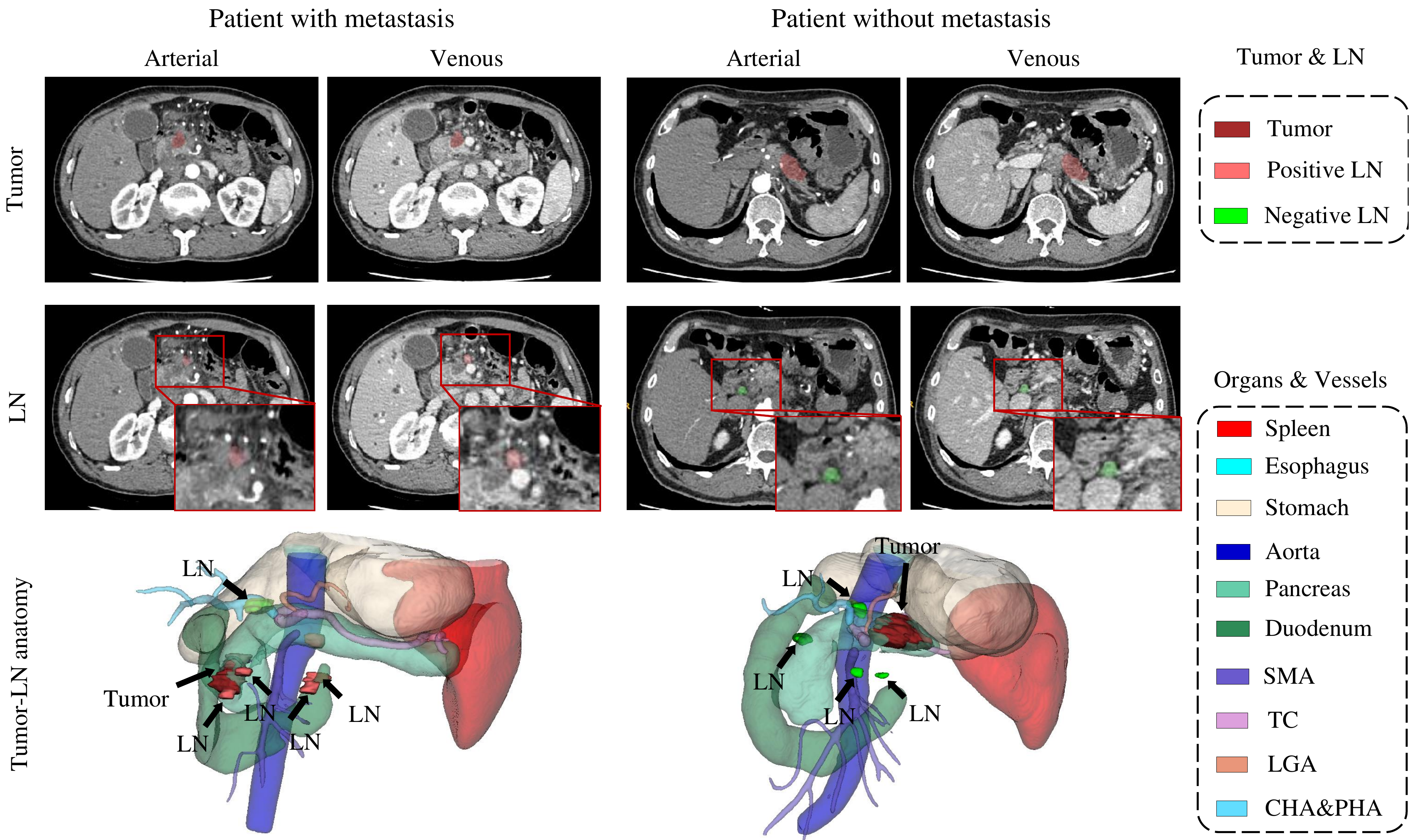}}
\caption{\small A visualization of pancreatic tumor (in dark red) and LNs (in pink red for positives or green for negatives) in multi-phase CT images and their spatial distributions corresponding to key anatomical structures as follows. SMA: superior mesenteric artery. TC$\&$SA: truncus coeliacus and splenic artery. LGA: left gastric artery; CHA$\&$PHA: common hepatic artery and proper hepatic artery.}
\label{Fig:distributionExample}
\end{figure*}

In this paper, we tackle the LN metastasis status prediction problem in patients with PDAC by first segmenting and identifying instances of LNs and then classifying the patients into {\it metastasis-positive} or {\it -negative} group. LNs are tiny structures that anatomically locate surrounding organs and vessels. Their locations have been mapped into 18 stations that are relevant for pancreatic cancer tumor according to their relative positions against adjacent anatomical structures, as defined by Japan Pancreas Society (JPS)\cite{kanehara2017classification} (see Supplementary Table 1 in the supplementary material for details). Examples of their spatial distribution are shown in Fig. \ref{Fig:distributionExample}. Metastasis happens when cancer cells spread from the primary tumor to LNs, causing enlargement of LNs and other underlying changes. Response 
Evaluation Criteria in Solid Tumors (RECIST) criteria \cite{eisenhauer2009new} defines the criteria for LN metastasis suspicion, i.e., nodes with short axis greater than 10{\it mm}, heterogeneity and central necrosis. However, these criteria are not pathognomonic since there exist false negatives associated with small node micrometastases and false positives with inflammatory nodes larger than 10{\it mm} in short axis. Hence, finding LNs in CT images is quite time-consuming and can be inconsistent depending on radiologists' subjective experiences. It is ambiguous for radiologists to identify nodal positivity accurately from CT without referring to pathology reports. The gold standard for determination of metastasis is based on post-operative pathological evaluation of pancreatectomy specimens. Automated yet reliable pre-operative LN segmentation and identification are highly desirable for patient surgical or RT treatment planing. 

LN segmentation is inherently challenging due to two reasons: 1) small to tiny sizes of LN cause extreme foreground-background class imbalance problem; 2) LNs have CT attenuation values similar to vessels and other soft-tissue structures, 
resulting in visual confusions. Existing work \cite{oda2018dense, bouget2019semantic, guo2021deepstationing} mainly adopt U-Net based deep networks \cite{ronneberger2015u, cciccek20163d,isensee2021nnu} as strong backbones, in which skip connections aggregate multi-level semantic representation and alleviate vanishing gradient problem. 
They incorporate anatomical context by directly taking organs$\&$vessels segmentation masks as either supervision targets \cite{oda2018dense} or additional inputs \cite{bouget2021mediastinal, guo2021deepstationing}. Concerns are remained that the relationship between lymphatic anatomy and adjacent anatomical structures is not well explored. We address it by introducing a distance-guided attention map to fully utilize the spatial priors. In our segmentation framework, the LN attention map is obtained via a pre-defined mapping from distance maps of related organs/vessels that have been integrated into UNet-based backbone to control the segmentation network's spatial focus. It simultaneously assists in improving sample selection strategy that filters out non-informative negative samples (called "informative negative selection") to tackle the class imbalance problem. The segmented LNs are labeled as positive/negative using radiologist's judgement as the standard that combines information from pathological results and CT intensities. A classification network is subsequently derived by sharing the same backbone with segmentation and initialized with the trained segmentation parameters. This strategy benefits the classification task from densely structured prediction in segmentation. By predicting LN metastasis in patients with PDAC, we employ a modified ResNet\cite{he2016deep} classification model. Tumor characteristics are proven to be important cues for metastasis \cite{li2020contrast,bian2019relationship,gao2020radiomics}, so we integrate both tumor and LN cues by taking as inputs the image patches of tumor and the patient-wise aggregation of LN segmentation and identification.

Our main contribution is four folds: 1) To the best of our knowledge, this work is the first to directly incorporate automated LN segmentation and identification for assisting preoperative LN metastasis status prediction for patients with PDAC. 2) We propose an attention-based LN segmentation network with the guidance of distances to nearby anatomical structures, explicitly exploiting the spatial priors and simultaneously addressing the foreground-background imbalance issue. 3) We design a compositive LN metastasis status prediction network combining both tumor and positive LN characteristics, showing the potential of tumor-LN guided cues. 4) Extensive quantitative experiments are conducted to evidently validate the effectiveness of our deep local attention network in both tasks of LN segmentation and LN metastasis status prediction, and external multi-center clinical evaluation is performed to demonstrate the generalization ability. Without loss of generality, our proposed method is applicable for finding the preoperative LN metastasis status of other types of solid tumor or cancers, such as liver or gastric cancers.

\section{Related Work}
\subsection{Lymph Node Segmentation}
Automated LN segmentation in CT images is an essential yet challenging task in medical image analysis. Traditional approaches tackle this problem by the means of atlas based search
space restriction\cite{feuerstein2012mediastinal}, spatial prior features combination\cite{liu2016mediastinal,liu2014mediastinal}, supervoxel clustering\cite{oda2017hessian}, etc. 
In recent years, U-Net based deep networks have shown remarkable performance in numerous organ or tumor segmentation tasks\cite{seo2019modified,huang20193d,kazemifar2018segmentation, oktay2018attention,gerard2019pulmonary}. nnUNet\cite{isensee2021nnu} further proposes a self-configuring approach, with automatic configurations including preprocessing,
network architecture, training and post-processing, that achieves robust performance and general applicability. To address the strong class imbalance issues in LN segmentation, four other anatomical structures are included as training targets \cite{oda2018dense} using 3D U-Net\cite{cciccek20163d} framework. \cite{bouget2019semantic} utilizes parallel networks of 2D U-Net\cite{ronneberger2015u} and Mask R-CNN\cite{he2017mask} with the supervision of all considered anatomical structures and LNs, benefiting from both semantic and instance segmentation. Another strategy to incorporate anatomical context is to take organ segmentation masks as additional channels of the input. \cite{bouget2021mediastinal} proposes an ensemble approach for
a slab-wise and a downsampled full volume based LN segmentation, taking the concatenation of CT image and segmented anatomical structure mask as input. DeepStationing\cite{guo2021deepstationing} presents a key referencing organ auto-search strategy and combines selected organs into the network via input concatentation for LN station parsing. All above methods implicitly exploit spatial priors of LNs by injecting the anatomical structure masks either as inputs or supervisions, hence the prior knowledge has not been fully exploited. More importantly, there is a lack of studies on how LN segmentation could be used for predicting metastasis.

\subsection{Lymph Node Metastasis Prediction}

\textbf{Radiomics Methods.}
Radiomics is a powerful technique for extracting quantitative image features with the purpose of clinical decision support, and thus widely used in cancer research  \cite{kumar2012radiomics,gillies2016radiomics,lambin2012radiomics,gao2020radiomics,meng20202d}. It converts imaging data into different types of hand-crafted features, including shape, intensity, texture and filter-based (e.g., wavelet, Laplacian of Gaussian) features. Applications of radiomics on predicting LN metastasis from primary tumor have been explored in many recent works\cite{ji2019biliary,wang2020ct,li2020contrast,bian2019relationship,yang2020integrating}. Radiomics features are first extracted from manually delineated tumor regions in any contrast-enhanced CT images. Feature selection and classification model construction (e.g., logistic regression, random forest) are then performed to give LN metastasis prediction for various cancer types like gastric cancer\cite{wang2020ct, meng20202d}, biliary tract cancer\cite{ji2019biliary} and PDAC\cite{li2020contrast,bian2019relationship,gao2020radiomics}. Relying only on primary tumor radiomics without considering LNs characteristics may limit the prediction performance, thus \cite{yang2020integrating} uses manual annotations of the largest LN visible in the gastric region and combines LN radiomics into the prediction model for gastric cancer. However, problem still remains because it simply involves the largest LN without identifying the nodal positivity.

\textbf{Deep Learning based Methods.} Recent advances in deep learning have made it a mainstream method of addressing the entire workflow of diagnosis and treatment for various cancer types on medical imaging, such as orapharageal cancer~\cite{ChengYao2021CCR}, lung cancer~\cite{xu2019CCR}, as well as pancreatic cancer~\cite{xia2021effective,zhao20213d,yao2020deepprognosis}. Deep neural networks are applied to LN metastasis in many studies\cite{zheng2020deep,dong2020deep,harmon2020multiresolution,jin2021deep}. In \cite{zheng2020deep}, deep features are extracted from tumor ROIs in bimodal image (i.e., US and SWE) using ResNet\cite{he2016deep}, and then fed into a SVM model for predicting axillary LN status in breast cancer. For gastric cancer, \cite{dong2020deep} combines DenseNet\cite{huang2017densely} features with some hand-crafted features, extracted from the 2D tumor ROI with the largest area in multi-phase CT images. To investigate metastasis in individual LN stations for gastric cancer, \cite{jin2021deep} develops a system of multiple independent ResNets with tumor ROIs and corresponding annotation masks as inputs where each ResNet is responsible to predict metastasis at one specific nodal station. Most existing studies capture only tumor characteristics for LN metastasis prediction, while the one leveraging LN radiomics requires manual delineation and considers simply the LN with the largest size \cite{yang2020integrating}. An automated and accurate process of LN segmentation and nodal positivity identification is hence of high importance for assisting metastasis prediction.

\section{Methodology}\label{sec1:methods}

\begin{figure*}[htb]
\centerline{\includegraphics[width=1\linewidth]{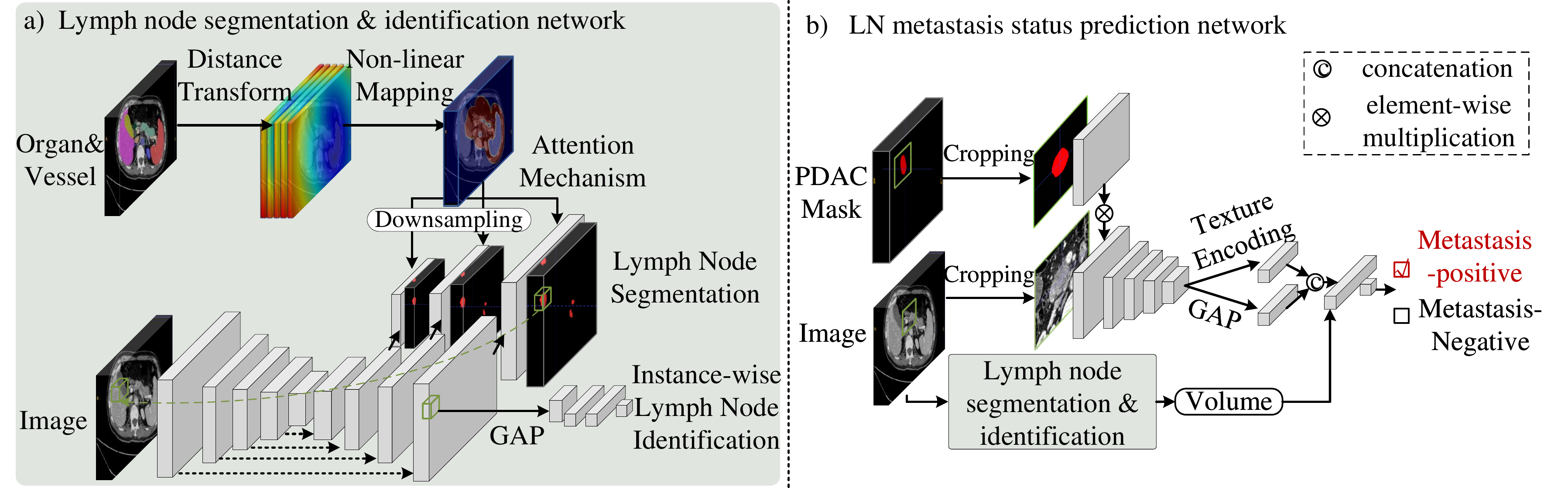}}
\caption{\small The proposed framework for (\textbf{a}) two-stage LN segmentation and identification and (\textbf{b}) LN metastasis status prediction .}
\label{Fig:framework}
\vspace{-2ex}
\end{figure*}

The overall framework is illustrated in Fig. \ref{Fig:framework}, which is composed of (a) distance-guided attention-based LN segmentation and identification network, and (b) tumor and LN combined metastasis status prediction network.

\subsection{Distance-guided Attention-based Lymph Node Segmentation and Identification Network} \label{Sec:LNSeg}
We perform LN detection from any input CT scan by a two-stage strategy: segmenting the image into two classes of LN and background voxels, followed by identifying segmented LN instances as positive or negative.

\subsubsection{Stage 1: Class-agnostic Lymph Node Segmentation}
Based on the spatial prior that LN stations are geometrically distributed or constrained around certain anatomical structures, we propose an attention based LN segmentation network by taking the distances to nearby organs/vessels into account. Our LN segmentation network differs from the strong baseline (i.e., nnUNet \cite{isensee2021nnu}) in that attention mechanism is applied to guide possible LN locations, with the advantage of reducing false positive predictions outside those locations. The intuition behind the attention module is that the attention map can cover regions adjacently constrained to those organs and vessels.
\begin{figure}[htb]
\centerline{\includegraphics[width=1\linewidth]{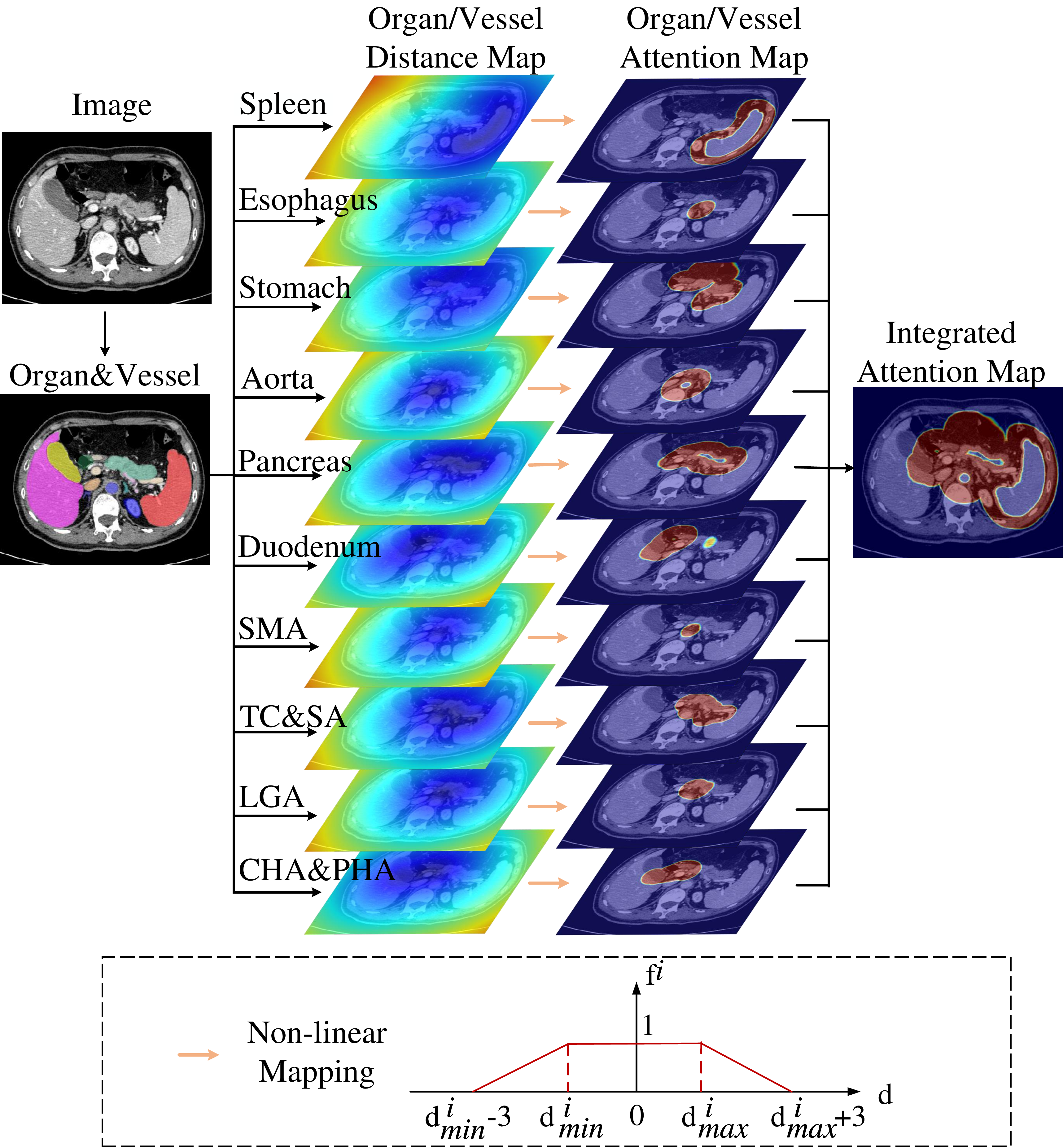}}
\caption{\small An illustration of attention map generation process.}
\label{Fig:AttmapGen}
\vspace{-2ex}
\end{figure}

\textbf{Attention Map Generation.} To explicitly capture and model the lymphatic anatomy, attention computation is implemented as a pre-defined geometric mapping function from organ$\&$vessel distance maps. An example of attention map generation process is shown in Fig. \ref{Fig:AttmapGen}. Specifically, given a multi-phase input CT volume $X \in \mathbb{R}^{N \times W \times H \times D}$, we first obtain organ$\&$vessel segmentation mask using nnUNet \cite{isensee2021nnu} model trained with 19 classes of annotations. Ten classes among them involved with 17 LN stations are used (see Supplementary Table 1 in the supplementary material for the definition of LN stations), i.e., spleen, esophagus, stomach, aorta, pancreas, duodenum, superior mesenteric artery (SMA), truncus coeliacus and splenic artery (TC$\&$SA), left gastric artery (LGA), common hepatic artery and proper hepatic artery (CHA$\&$PHA). Note that station 15\# (LNs along middle colic artery) is left aside here since it is related to distant metastasis that rarely happens in our patient population. A SDT is applied to each class of the segmentation mask $M \in \{0, 1,2,...,10\}^{W \times H \times D}$, generating a total of 10 organ/vessel distance maps $D^{i}$ where $i \in \{1, 2, ..., 10\}$ is the index of organ/vessel class. $D^{i}$ has positive values at the voxels outside the $i$-th organ/vessel and negative scores inside it. Intuitively, LNs are likely to appear within a certain range of distance to each organ/vessel, which requires paying attention to. To obtain the distance-guided attention maps, $D^{i}$ is passed to an isosceles trapezium-shaped non-linear mapping function (see Fig. \ref{Fig:AttmapGen}), formulated as 
\begin{equation} \label{Eq:mapping}
f^i (d)=\left\{
\begin{aligned}
&1,   &{d^i_{\mathrm{min}} \leq d \leq d^i_{\mathrm{\mathrm{max}}}}\\
& -\frac{(d-d^i_{\mathrm{max}}-3)}{3}, &{d^i_{\mathrm{max}} \textless d \textless d^i_{\mathrm{max}} + 3}\\
& \frac{(d-d^i_{\mathrm{min}}+3)}{3} , &{d^i_{\mathrm{min}} -3 \textless d \textless d^i_{\mathrm{min}}} \\
&0, &{Otherwise} 
\end{aligned}
\right..
\end{equation}
where $d$ is the individual element in $D^{i}$; $d^i_{\mathrm{min}}$ and $d^i_{\mathrm{max}}$ determine the distance range in {\it mm}; the smooth border 3{\it mm} is chosen empirically. This mapping function converts the distance maps to the attention scores ranging from 0 to 1, with 1 indicating possible locations of LNs, 0 indicating impossible locations, and decimals lying in between. The $i$-th attention map $A^{i}$ is obtained by $A^{i}=f^i (D^{i})$. 

The final attention map $A^{all}$ is produced by integrating all of the organ/vessel-specific attention maps, thus $A^{all}$ can cover the whole areas that need attending to. In specific, $A^{all}$ takes the element-wise maximum of all $A^{i}$ except for the voxels inside an organ/vessel, illustrated as 
\begin{equation} \label{Eq:a}
a^{all}_{\mathrm{v}}=\left\{
\begin{aligned}
& \max_{i=1,2,...,10} a^{i}_{\mathrm{v}},& m_{\mathrm{v}} = 0\\
&a^{i}_{\mathrm{v}}, & m_{\mathrm{v}} = i\\
\end{aligned}
\right..
\end{equation}
where $a^{*}_{\mathrm{v}}$ and $m_{\mathrm{v}}$ are the values of $A^{*}$ and $M$ at the voxel $\mathrm{v}$.

\textbf{Attention based Lymph Node Segmentation.} After obtaining the distance-guided attention map, we incorporate it to the segmentation network with 3D nnUNet\cite{isensee2021nnu} as the backbone. The attention mechanism emphasizes LN-related regions by spatially scaling the features with the attention map. Given the input image $X$ and the one-hot segmentation label $Y$, the deep features at the penultimate layer are extracted and multiplied element-wisely with $A^{all}$. It is finally passed through a convolution block with a softmax layer to produce the segmentation output $P \in \mathbb{R}^{2 \times W \times H \times D}$. 

Due to GPU memory limitation, patch-based training strategy is employed. nnUNet randomly samples 3D image patches from the whole CT scan and enforces that more than a third of samples in a batch contain at least one foreground class to control the foreground-to-background ratio. Considering the extreme class imbalance problem caused by the small LN targets, we improve it with the ``informative negative selection" scheme. Note that our proposed attention mechanism helps block out features at the voxels with a certain distance to or inside all organs and vessels, resulting in lots of non-informative negative patches filled with 0 by applying zero attention scores. Thus we can naturally throw out those 
non-informative patches, and select patches containing at least one attention score $\textgreater$ 0 (called informative patches) for training. This sampling strategy further boosts the network's concentration on targeted regions of interest (ROIs) surrounding organs/vessels.

For training objectives to better balance precision and recall , we modify the Dice loss in nnUNet with Tversky loss \cite{salehi2017tversky}:
\begin{equation} 
L_{T}= - \frac{2}{\lvert V\rvert } \frac{\sum_{\mathrm{v}} p_{1,\mathrm{v}} y_{1,\mathrm{v}}}{2\sum_{\mathrm{v}} p_{1,\mathrm{v}} y_{1,\mathrm{v}} + \alpha \sum_{\mathrm{v}} p_{1,\mathrm{v}} y_{0,\mathrm{v}}  + \beta \sum_{\mathrm{v}} p_{0,\mathrm{v}} y_{1,\mathrm{v}} }
\end{equation}
where $\lvert V\rvert$ is the number of voxels. $p_{1,\mathrm{v}}$ is the probability of voxel $\mathrm{v}$ being a LN, and $p_{0,\mathrm{v}}$ is the probability being a non-LN. Also, $y_{1,\mathrm{v}}$ is 1 for a LN voxel and 0 for a non-LN voxel, and vice versa for the $y_{0,\mathrm{v}}$. In practice, we set $\alpha=0.5$ and $\beta=1.5$ to emphasis on false negatives and boost recall. The whole network is trained by the combination of cross entropy loss $\mathcal{L}_{\text{CE}}$ and Tversky loss $\mathcal{L}_{\text{T}}$ with equal weights as in nnUNet.
\begin{align}
&\mathcal{L}_{\text{CE}} = - \frac{1}{\lvert V\rvert} \sum_{\mathrm{v}} \sum_{\mathrm{k}=0,1} y_{k,\mathrm{v}} \log(p_{k,\mathrm{v}}) \\ 
&\mathcal{L}_{\text{SEG}} = \mathcal{L}_{\text{CE}} + \mathcal{L}_{\text{T}}
\vspace{-1em}
\end{align}
Following nnUNet, the network is trained with deep supervision, i.e., losses are calculated over multi-resolution outputs given by final and intermediate layers of the decoder, and the corresponding downsampled ground-truth (GT) masks are used as targets. Here attention mechanism is applied in a multi-scale manner. That is, the attention map, after downsampled to match the resolution, is injected to the intermediate decoder feature for each deep supervision output.

\subsubsection{Stage 2: Instance-wise Lymph Node Identification}
After segmenting LN instances from the whole CT image, we then classify them into either positive or negative class. To benefit from the already trained dense segmentation network of stage 1, the task of LN instance identification reuses 3D nnUNet backbone and is initialized using the trained segmentation parameters, with a new classification head added upon it.
Cross entropy loss is adopted to finetune the whole network for classifying the instance as positive/negative. To generate LN instances, we crop patches centered at the connected components of the segmentation mask. GT LN instances are cropped and employed in the training phase. While at inference time, we can apply the classification network to identify each segmented LN of stage 1, and obtain a class-aware LN segmentation mask.

\subsection{Tumor and Lymph Node Combined LN Metastasis Status Prediction Network}
Besides LNs themselves, imaging characteristics in the primary tumor play an important role in predicting the status of LN metastasis. To further boost the performance, we build a combined classification network, integrating both PDAC and LNs related information. In contrast to previous work that only consider tumor characteristics\cite{li2020contrast,bian2019relationship,gao2020radiomics}, our method benefits from directly observing the status of LN instances by automated LN segmentation and identification.

Given a CT image and the corresponding PDAC mask, 2D slices with the top three largest PDAC areas 
in each of axial, sagittal, and coronal planes are cropped, resulting in nine image patches in total. Each image patch is fed into a ResNet~\cite{he2016deep} pre-trained on ImageNet~\cite{deng2009imagenet} for metastasis prediction. Inspired by \cite{eppel2018classifying}, a side branch with the PDAC mask as input is added to encourage the network to concentrate on the PDAC region. Specifically, the side branch consists of a Conv-ReLU block and maps the input mask to a feature map with the same shape as the output of ``Conv1" layer in ResNet. It is then integrated into the backbone by element-wise multiplication with the ``Conv1" feature. Our initial  experiment empirically shows that such incorporation produces better performance than direct input-level fusion, as the convolution in the side branch learns which region to focus on in each channel of ``Conv1" feature (e.g., regions inside the mask, around the mask border or outside the mask).  To be better aligned with the pre-trained backbone and eliminate the initial effect of the side branch, the weights and biases in the convolution layer are initialized to 0 and 1 respectively. Before classification, we additionally employ a Texture Encoding Layer (TEL) \cite{zhang2017deep} on top of the ``Layer4" feature $\mathcal{F}_{L4}$ to extract respective texture representation. TEL serves as an orderless feature pooling layer that encodes spatially invariant representation describing the feature distributions, which benefits texture recognition of the PDAC region. The original deep feature is merged with the texture feature to form an enhanced representation $\mathcal{F}$:
\begin{align}
&\mathcal{F} = \mathrm{Concat}(\mathrm{GAP}(\mathcal{F}_{L4}), \mathrm{TEL}(\mathcal{F}_{L4}))
\vspace{-1em}
\end{align}
where Concat and GAP denote feature concatenation and global average layer, respectively.

We further integrate LN-related cues into the network given the LN segmentation and identification results described in Section \ref{Sec:LNSeg}. A patient is considered as metastasis-positive if there exists at least one positive LN, thus it is very sensitive to the false positives in LN identification. Therefore, we employ the volume of positive LN as the feature instead of 
its binary status of presence/absence, based on the fact that positive LNs tend to have larger volume than negative ones. The volume of the largest positive LN in each patient $\mathcal{V}_{\mathrm{pLN}}^{\max}=\max \limits_{ln \in \{positive\ LNs\}} \mathcal{V}_{ln}$ (in {\it mm$^3$}) is mapped to a vector-shaped feature, and fused with $\mathcal{F}$ by element-wise addition, formulated as follows:
\begin{align}
&\mathcal{F}_{\mathrm{comb}} = \mathrm{FC}(\mathrm{BN}(\mathcal{V}_{\mathrm{pLN}}^{\max})) + \mathcal{F}
\vspace{-1em}
\end{align}
where $\mathrm{FC}$ and $\mathrm{BN}$ denote the full-connected layer and batch normalization layer, respectively. Other LN features, such as the average or total volume of positive LNs, are also evaluated, with the current setting giving the best result. Finally, the classification probabilities generated from nine image patches are averaged to given an ensembled prediction for a patient.

\section{Experiments}
In this section, we first demonstrate the multicenter datasets (i.e. the discovery dataset and two external datasets) and implementation details, and then elaborate the strategy we use to generate PDAC segmentation masks. Subsequently we present results on the discovery dataset in each step of our method, including organ$\&$vessel segmentation, attention map generation, LN segmentation and identification, and LN metastasis status prediction. Finally, external validation is conducted to evaluate the generalization performance of LN metastasis status prediction, with only pathology reports accessible in two external datasets.

\subsection{Experimental Settings}

\subsubsection{Dataset}
\begin{table*}[htb]
\small
\centering
\caption{\small Demographic distributions and tumor characteristics in the three datasets (\textbf{Discovery dataset}, \textbf{Ext-validation dataset 1} and \textbf{Ext-validation dataset 2}). Median [interquartile range, 25th–75th percentile] values are reported for continuous variables.}
\label{Tab:Dataset}
\begin{tabular}{lccc}
\toprule
\multirow{2}{*}{Characteristics} & Discovery dataset  & Ext-validation dataset 1& Ext-validation dataset 2 \\
& (n=749) &  (n=132) & (n=59) \\
\midrule
Gender, n ($\%$) & & &\\
Female & 282 (38$\%$)& 60 (45$\%$)& 28 (47$\%$)\\
Male & 467 (62$\%$) & 72 (55 $\%$)& 31 (53$\%$)\\
Age at Diagnosis, yrs & 63 [56-69]& 60 [53-65]& 58 [51-62]\\
pT Stage, n ($\%$) & & &\\
pT1 / pT2 & 92 (12$\%$) / 314 (42$\%$)& 24 (18$\%$)/ 80 (61$\%$)& 10 (17$\%$) / 31 (53$\%$)\\
pT3 / pT4 & 316 (42$\%$) / 13 (2$\%$) & 15 (11$\%$)/ 13 (10$\%$)& 5 (8$\%$)/ 13 (22$\%$)\\
Missing & 14 (2$\%$) & 0 (0 $\%$) & 0 (0 $\%$)\\
pN Stage, n ($\%$) & & &\\
pN0 & 398 (53$\%$)& 93 (70$\%$)& 22 (37$\%$)\\
pN1& 242 (32$\%$) & 32 (24$\%$)& 22 (37$\%$)\\
pN2 & 109 (15$\%$)& 7 (5$\%$)& 15 (25$\%$)\\
Tumor Size, cm & 3.0 [2.5-4.1]& 2.7 [2.2-3.0]& 2.9 [2.2-3.4]\\
Tumor Location, n ($\%$) & & &\\
Head / Uncinate & 475 (63$\%$)& 56 (42$\%$)/ 52 (39$\%$)& 35 (59$\%$)/ 22 (37$\%$)\\
Body / Tail & 274 (37$\%$)& 2 (2$\%$)/ 22 (17$\%$)& 2 (3$\%$)/ 0 (0$\%$)\\
Positive LN Volume, {\it mm}$^3$& 665[210-804]& - & -\\
Negative LN Volume, {\it mm}$^3$& 300[106-377]& - & -\\
\bottomrule
\end{tabular}
\vspace{-1em}
\end{table*}
We conduct a multicenter study on three independent datasets with a total of 940 patients collected from Changhai Hospital in Shanghai, Shengjing Hospital in Liaoning Province, and Tianjin Cancer Hospital in Tianjin. All patients had a pathologically confirmed diagnosis of PDAC, and contrast-enhanced CT scans of arterial (A) and venous (V) phases acquired before treatment were included in this study. We labeled LNs on the dataset from Changhai Hospital (denoted as Discovery dataset), and developed our model on it using nested cross-validation (CV). The rest two datasets from Shengjing Hospital and Tianjin Cancer Hospital (denoted as Ext-validation dataset 1 and Ext-validation dataset 2) were used as external validation sets with only pathologically diagnosed LN metastasis status provided. This study was reviewed and approved by the Biomedical Research Ethics Committee of the institution (No. CHEC2021164), and was performed in accordance with the ethical standards of the 1964 Declaration of Helsinki. The requirement for patient informed consent was waived by the Institutional Review Board due to the retrospective nature of the study and because all procedures performed were part of routine care.

\textbf{Discovery dataset}  contains CT scans of 749 patients, among which there are 351 positive samples (patients with LN metastasis) and 398 negative samples (patients without LN metastasis). 
The annotation of LNs was performed by two board-certified radiologists (XF with 7 and MZ with 5 years of experiences in pancreatic imaging) with referring to pathology report under supervision of a senior radiologist (YB) with 17 years of experiences in pancreatic imaging.  There are totally 2,467 labeled LNs, of which 476 are positive and the rest are negative. In specific, 351 metastasis-positive patients contain 476 labeled positive and 322 labeled negative LNs, and 398 metastasis-negative patients have the rest 1,669 labeled negative LNs. 

This dataset was split using nested five-fold CV, with 64\%, 16\% and 20\% as training, validation and testing sets in each CV round. As for the primary tumor, 163 patients among the whole dataset were annotated with 3D tumor masks by two radiologists (XF and MZ). We use these 163 patients as the testing set and the remaining unlabeled 586 patients as the training set for an annotation-efficient PDAC segmentation. Additionally, we generate pseudo annotations of 17 classes of organs and vessels using the self-learning segmentation model described in \cite{yao2021deepprognosis}, and manually annotate extra two vessels (LGA, CHA$\&$PHA) and extend other two vessels (SMA and TC$\&$SA) under the supervision of a radiologist (XF) for 50 patients randomly sampled from our dataset. 40/10 of these patients are used as training and validation sets for organ$\&$vessel segmentation, respectively. 

\textbf{Ext-validation dataset 1} contains CT scans of 132 patients with 39 positive and  93 negative patients; \textbf{Ext-validation dataset 2} contains 59 patients with 37 positive and 22 negative patients. More detailed information on three datasets can be seen in Table \ref{Tab:Dataset}.

\subsubsection{Implementation Details}
In our experiments, CT images of arterial phase are registered to venous phase using DEEDS\cite{heinrich2013towards}, and they are all resampled to a median spacing of 0.68 $\times$ 0.68 $\times$ 0.80 mm. For LN segmentation and organ$\&$vessel segmentation, sub-volumes of  160 $\times$ 192 $\times$ 80 are randomly cropped as training patches. In the non-linear mapping from distance maps to attention maps for our LN segmentation, parameters of the mapping function are determined by grouping GT LN voxels according to which organ/vessel is closest to, and calculating the minimum and maximum distances to organ/vessel boundaries in each group. Parameters are listed in Table \ref{Tab:mappingPara}, in which negative values indicate voxels inside organ/vessel.
\begin{table}[htbp]
\small
\centering
\caption{Parameters (i.e. $\mathrm{d_{min}}$ and $\mathrm{d_{max}}$) of non-linear mapping function for each organ or vessel. SMA: superior mesenteric artery. TC$\&$SA: truncus coeliacus and splenic artery. LGA: left gastric artery; CHA$\&$PHA: common hepatic artery and proper hepatic artery.}
\label{Tab:mappingPara}
\begin{tabular}{lcc}
\toprule
{Organ/Vessel} & $\mathrm{d_{min}}$ (mm)& $\mathrm{d_{max}}$ (mm)\\
\midrule
Spleen & 0 & 16\\
Esophagus & 0& 25\\
Stomach & -2& 18\\
Aorta &0& 28\\
Pancreas & -5& 20\\
Duodenum & -5& 22\\
SMA & -1& 20\\
TC$\&$SA & -2& 18\\
LGA  & 0 & 21\\
CHA$\&$PHA  & 0 & 20 \\
\bottomrule
\end{tabular}
\vspace{-1em}
\end{table}

As for instance-wise LN identification, 3D image training samples are generated by cropping a 96 $\times$ 96 $\times$ 80 sub-volume centered per each GT LN. SGD optimizer with Nesterov momentum ($\mu$ = 0.95) is adopted to train the network, whose initial learning rate and weight decay are $5 \times 10^{-4}$ and $1 \times 10^{-4}$, respectively. Furthermore, the final LN metastasis status prediction model takes 2D inputs of 224 $\times$ 224 centered at PDAC, and is trained using the same optimizer as above. Details of the network architecture are presented in the supplementary material.

\subsection{PDAC Segmentation Mask Acquisition/Harvesting}
We employ an annotation-efficient strategy to generate 3D masks of tumors for the labor cost reduction purpose. Specifically, we start with the PDAC segmentation model trained with arterial-late phase described in \cite{zhao20213d} to generate pseudo annotations. Next, the model is fine-tuned under the supervision of pseudo annotations and then applied to produce segmentation masks on our dataset. To obtain the PDAC segmentation model on venous phase, those segmentation masks are registered to venous phase and are then used to train a nnUNet model from scratch to generate the final 3D masks of tumors. We evaluate the final PDAC segmentation model on the labeled testing set. Median Dice score, average surface distance (ASD, {\it mm}), and Hausdorff distance (HD, {\it mm}) are 0.683, 2.186, and 12.805 respectively.

\subsection{Evaluation of Organ$\&$Vessel Segmentation and Attention Maps}
To evaluate the performance of organ$\&$vessel segmentation, a testing set of 19 randomly selected CT volumes with ten classes of organ/vessel is manually annotated by a radiologist (XF). To reduce the annotation burden, all dense CT volumes are downsampled to 5{\it mm} in the slice thickness dimension. We compare our self-learning model with the pseudo annotation generator \cite{yao2021deepprognosis}, which is able to segment eight of ten classes (except for LGA and CHA$\&$PHA) on single-phase CT. Dice score, ASD ({\it mm}), and HD ({\it mm}) are adopted as the evaluation metrics and the results are provided in Table \ref{Tab:ovseg}. Our model that is trained on two phases outperforms \cite{yao2021deepprognosis} on seven of eight organs/vessels. Note that SMA and TC$\&$SA masks segmented by \cite{yao2021deepprognosis} contain shorter parts compared with those segmented by our model, therefore, resulting in significantly lower performance than ours (0.331 lower Dice score in SMA, and 0.171 lower in TC$\&$SA). 

\begin{table}[htb]
\small
\centering
\caption{\small Quantitative Performance of Organ$\&$Vessel Segmentation. A: arterial. V: venous. SMA: superior mesenteric artery. TC$\&$SA: truncus coeliacus and splenic artery. LGA: left gastric artery; CHA$\&$PHA: common hepatic artery and proper hepatic artery. }
\label{Tab:ovseg}
\resizebox{0.5\textwidth}{!}{
\begin{tabular}{llcccc}

\toprule
{Organ/Vessel} &Methods&CT Phases&Dice & ASD ({\it mm}) & HD ({\it mm})\\
\midrule
{Spleen } & \cite{yao2021deepprognosis} & A & 0.938 & 0.643 & 14.252 \\
& \cite{yao2021deepprognosis} & V & 0.954 & 0.422 & 11.107 \\ 
& Ours & A+V& \textbf{0.959}& \textbf{0.384}& \textbf{8.129}\\

{Esophagus} &\cite{yao2021deepprognosis} & A & 0.557 & 0.936 & 13.897 \\
& \cite{yao2021deepprognosis} & V & 0.598 & 0.854 & 11.003 \\
& Ours& A+V&\textbf{0.745} & \textbf{0.641}& \textbf{8.125}\\

{Stomach} &\cite{yao2021deepprognosis} & A & 0.846 & 2.223 & 35.338 \\
& \cite{yao2021deepprognosis} & V & 0.813 & 3.765 & 43.114 \\
& Ours& A+V& \textbf{0.905}& \textbf{1.519}& \textbf{19.183}\\

{Aorta } &\cite{yao2021deepprognosis} & A & 0.893 & 0.519 & 8.130 \\
& \cite{yao2021deepprognosis} & V & \textbf{0.924} & 0.417 & 6.158 \\
& Ours& A+V& 0.920 & \textbf{0.359}& \textbf{5.863}\\

{Pancreas } &\cite{yao2021deepprognosis} & A & 0.712 & 2.905 & 25.880 \\
& \cite{yao2021deepprognosis} & V & 0.756 & 1.897 & 19.258 \\
& Ours& A+V& \textbf{0.847}& \textbf{0.975}& \textbf{12.859}\\

{Duodenum } &\cite{yao2021deepprognosis} & A & 0.613 & 2.976 & 34.187 \\
& \cite{yao2021deepprognosis} & V & 0.665 & 3.366 & 32.174 \\
& Ours& A+V&\textbf{ 0.764}& \textbf{1.892}& \textbf{29.131}\\

{SMA } &\cite{yao2021deepprognosis} & A & 0.387 & \textbf{0.663} & 68.869 \\
& \cite{yao2021deepprognosis} & V & 0.415 & 0.710 & 68.098 \\
& Ours& A+V& \textbf{0.746}& 0.860& \textbf{28.840}\\

{TC$\&$SA} & \cite{yao2021deepprognosis} & A & 0.563 & 0.780 & 43.974 \\
& \cite{yao2021deepprognosis} & V &  0.407 & 1.245 & 51.432 \\
& Ours & A+V&\textbf{ 0.734}& \textbf{0.305}& \textbf{22.224}\\

{LGA } & \cite{yao2021deepprognosis} & A & - & - & - \\
& \cite{yao2021deepprognosis} & V & - & - & - \\ 
& Ours& A+V& 0.651& 0.371& 10.420\\

{CHA$\&$PHA } &\cite{yao2021deepprognosis} & A &  - & - & - \\
& \cite{yao2021deepprognosis} & V &  - & - & - \\
& Ours & A+V& 0.715& 1.424& 24.239\\
\bottomrule
\end{tabular}
}
\vspace{-1em}
\end{table}

Qualitative evaluation of organ$\&$vessel segmentation examples as well as the corresponding attention maps are visualized in Fig. \ref{Fig:LNSeg} (b).

\subsection{Evaluation of Lymph Node Instance Segmentation and Identification}

\begin{table}[htb]
\small
\centering
\caption{\small Average instance-wise LN classification performance across 5 folds. The results are reported on GT instances.}
\label{Tab:LNCls_Results}

\begin{tabular}{lc}
\toprule
{Metric} & Performance \\
\midrule
AUC  & 0.854  \\
Accuracy  & 0.789 \\
Balanced accuracy & 0.771 \\
Sensitivity & 0.742 \\
Specificity & 0.800 \\
\bottomrule
\end{tabular}
\vspace{-1em}
\end{table}

\begin{table*}[htb]
\small
\centering
\caption{\small Performance comparison on LN segmentation before and after instance-wise identification (denoted as \emph{Class-agnostic Seg} and \emph{Class-aware Seg}). Pos and Neg denote positive and negative LNs. Results are averaged across 5 folds. Wilcoxon signed rank test is conducted on voxel-wise Dice and instance-wise F-measure. * indicates $p$-value $\textless$ 0.05. NS indicates no significance.}
\label{Tab:LNSeg_Results}
\begin{tabular}{lcccccccc}
\toprule
\multirow{2}{*}{Stage} & \multirow{2}{*}{Class} &\multirow{2}{*}{Method} &  \multicolumn{3}{c}{Voxel-wise Metrics ($\%$)} & \multicolumn{3}{c}{Instance-wise Metrics ($\%$)}  \\
\cmidrule{4-6} \cmidrule(lr){7-9} 
& & & Dice & Recall & Precision & F-measure & Recall &  Precision\\ 
\midrule
{Class-agnostic Seg}& \multirow{2}{*}{-} &\footnotesize{nnUNet} & 45.9$^ *$ & 75.4 &  36.2 & 36.1$^*$ &  \textbf{81.0} &25.3\\
(before identification)& & Ours & \textbf{47.7}$^{ref}$ & \textbf{77.7} & \textbf{37.5} & \textbf{40.6}$^{ref}$ & 80.9 & \textbf{29.9}\\
\cmidrule(lr){1-9}
& \multirow{2}{*}{Pos} &\footnotesize{nnUNet} & 10.2$^*$& 32.3& 11.3& 11.7$^*$& 36.1& 12.0\\
Class-aware Seg & & Ours & \textbf{12.0}$^{ref}$& \textbf{38.9}& \textbf{11.7}& \textbf{13.5}$^{ref}$& \textbf{41.5}& \textbf{13.3}\\
\cmidrule(lr){2-9}
(after identification) & \multirow{2}{*}{Neg} & \footnotesize{nnUNet} & 27.5$^{NS}$& \textbf{51.1}& 25.4& 27.7$^*$& \textbf{60.0}& 22.9\\
& &Ours & \textbf{27.7}$^{ref}$& 49.0& \textbf{27.0}& \textbf{28.9}$^{ref}$& 56.2& \textbf{25.8}\\
\bottomrule
\end{tabular}
\vspace{-0.6em}
\end{table*}

\begin{figure*}[!h]
\begin{center}
\includegraphics[width=0.7\linewidth]{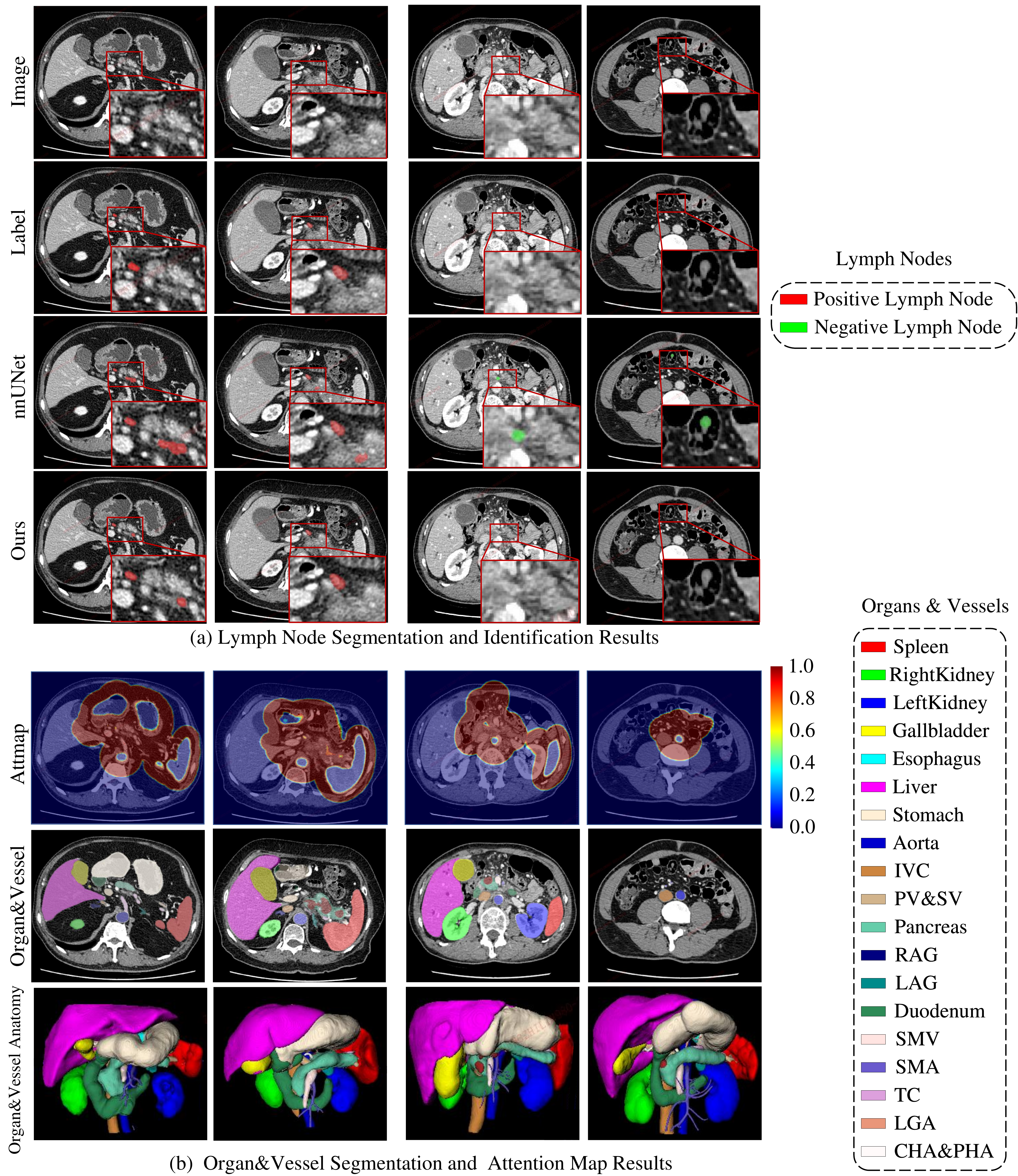}
\end{center}
\vspace{-2ex}
\caption{\small Examples of (a) LN segmentation and identification results, and (b) Organ$\&$Vessel segmentation and attention map results. }
\vspace{-0.2em}
\label{Fig:LNSeg}
\end{figure*}

\textbf{Quantitative Evaluation.} LNs are first detected by the class-agnostic segmentation model, and then identified as positive/negative by applying the classification model on the cropped instances. For positive/negative LN identification, our classification model is trained with Ground-Truth (GT) LNs, yielding an average AUC of 0.854 across 5 folds (in Table \ref{Tab:LNCls_Results}). At inference, the automatically segmented LNs are cropped and then identified by the classification model. To evaluate the segmentation performance before and after identification, we compare our method with a strong baseline, nnUNet \cite{isensee2021nnu}. 
The segmentation accuracy is measured by voxel-wise metrics (i.e., Dice, Recall, Precision) and instance-wise metrics (F-measure, Recall, Precision). To achieve the statistical analysis, we apply 1,000 iterations of Wilcoxon signed rank test to voxel-wise Dice and instance-wise F-measure. Results are provided in Table \ref{Tab:LNSeg_Results}. An instance is considered successfully detected if its (intersect-over-union) IoU score between the segmentation mask and GT mask is $\ge$ 30 $\%$. Before identification, our segmentation model significantly outperforms nnUNet on both voxel-wise and instance-wise metrics, with the voxel-wise Dice increasing from 45.9$\%$ to 47.7$\%$ and the instance-wise F-measure increasing from 36.1$\%$ to 40.6$\%$, as shown in Table \ref{Tab:LNSeg_Results}. In addition, our model also yields superior performance in terms of both positive and negative LNs after identification, achieving 1.8$\%$ higher voxel-wise Dice and 1.8$\%$ higher instance-wise F-measure in terms of positive LNs, and 0.2$\%$ higher voxel-wise Dice and 1.2$\%$ higher instance-wise F-measure in terms of negative LNs. In total five out of six comparisons, our method is statistically significantly better or more accurate (i.e., with $p$-value $\textless$ 0.05) in LN segmentation than the nnUNet baseline (implemented without the attention maps). 

\begin{table*}[!ht]
\renewcommand\arraystretch{0.95}
\small
\centering
\caption{\small  Performance comparison and ablation study on LN metastasis status prediction of Discovery dataset. Results are averaged across 5 folds. Wilcoxon signed rank test is conducted on balanced accuracy. * indicates $p$-value $\textless$ 0.05. NS indicates no significance.}
\label{Tab:LNMAbaltionBaseline}
\begin{tabular}{llllll}

\toprule
\multirow{2}{*}{Method} & Balanced Accuracy & AUC & Accuracy &  Sensitivity & Specificity \\
& [95$\%$ CI] & [95$\%$ CI] & [95$\%$ CI] & [95$\%$ CI] & [95$\%$ CI] \\
\midrule
\multirow{2}{*}{CT-reported LN status} & 0.599$^*$ & - & 0.599 & 0.588 & 0.609 \\
& [0.564-0.634] & & [0.565-0.634] &  [0.538-0.635] & [0.558-0.657] \\
\multirow{2}{*}{Radiomics } & 0.597$^*$ & 0.648 & 0.603 &  0.508 & 0.686\\
& [0.563-0.633] & [0.598-0.681] & [0.569-0.637] &  [0.456-0.561] & [0.638-0.734] \\
{Radiomics + } & 0.604$^*$& 0.654& 0.610&  0.524& 0.684\\
{CT-reported LN status} & [0.572-0.641] & [0.612-0.692] & [0.575-0.644] &  [0.470-0.581] & [0.641-0.731] \\
\multirow{2}{*}{ResNet3D} & 0.562$^*$& 0.609& 0.554&  0.599& 0.524\\
& [0.521-0.593] & [0.550-0.631] & [0.519-0.587] &  [0.538-0.644] & [0.475-0.568] \\
\multirow{2}{*}{ResNet2D} & 0.571$^*$& 0.631& 0.574&  0.568& 0.574\\
& [0.540-0.609] & [0.590-0.667] & [0.537-0.607] &  [0.519-0.624] & [0.530-0.628] \\
\multirow{2}{*}{DeepTEN} & 0.588$^*$& 0.634& 0.593&  0.609& 0.566 \\
& [0.560-0.628] & [0.599-0.679] & [0.559-0.628] &  [0.564-0.667] & [0.520-0.621] \\
\midrule
\multirow{2}{*}{ClsfromPDAC} & 0.599$^*$& 0.654 & 0.597 &  0.600& 0.597\\
&[0.558-0.634]& [0.608-0.685] & [0.561-0.633] & [0.547-0.647]&[0.550-0.646]\\
\multirow{2}{*}{ClsbyLNSeg w/o Attn} & 0.545$^*$& 0.590 & 0.566&  0.433& 0.657\\
& [0.525-0.593] & [0.548-0.625] & [0.534-0.601] &  [0.393-0.499] & [0.623-0.716] \\
\multirow{2}{*}{ClsbyLNSeg w/ Attn} & 0.563$^*$& 0.603& 0.572&  0.351& \textbf{0.775}\\
& [0.530-0.594] &[0.564-0.642] & [0.539-0.605] &  [0.299-0.396] & [0.731-0.814] \\
{Ours (ClsfromPDAC + } & \textbf{0.633}$^{ref}$ & \textbf{0.682} & \textbf{0.635} & \textbf{0.618}& 0.649\\
{ClsbyLNSeg w/ Attn)}& [0.599-0.669] &[0.640-0.717] & [0.601-0.669] &  [0.567-0.664] & [0.603-696] \\

\bottomrule
\end{tabular}
\vspace{-1.0em}
\end{table*}

\begin{figure*}[!ht]\centering
    \subfloat[\label{fig:a}]{ \includegraphics[width=0.48\textwidth]{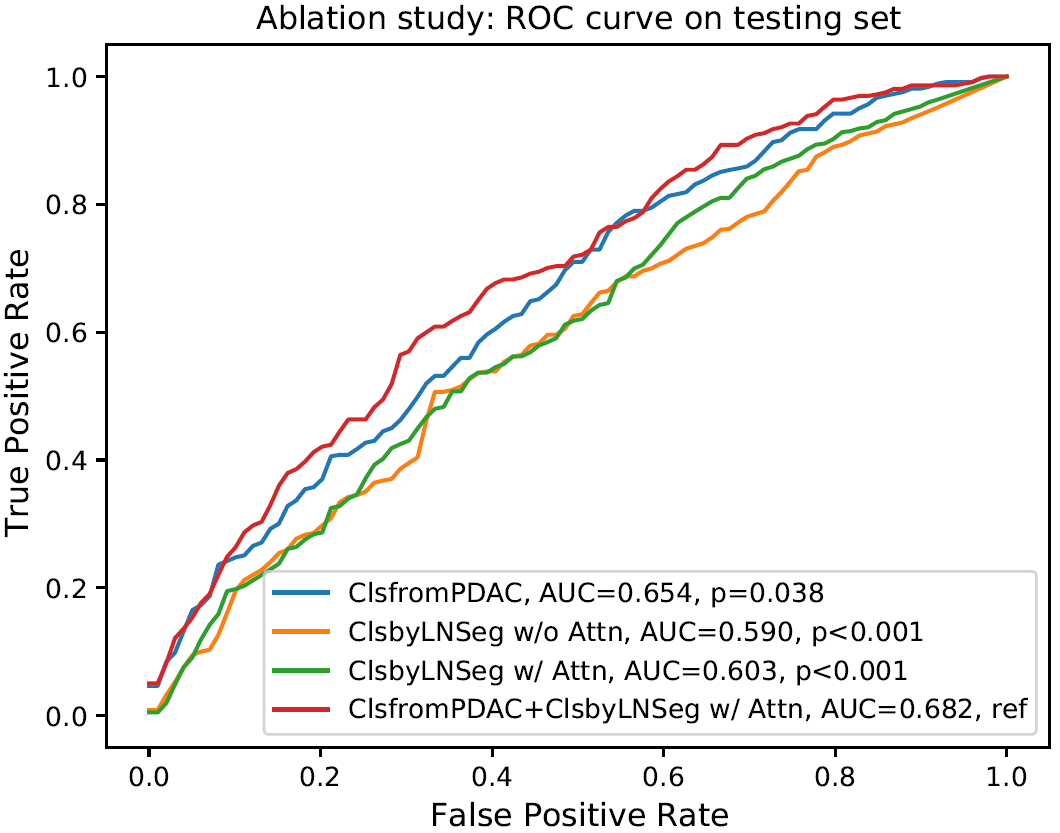}\label{Fig:AblationROC}}
    \subfloat[\label{fig:a}]{ \includegraphics[width=0.48\textwidth]{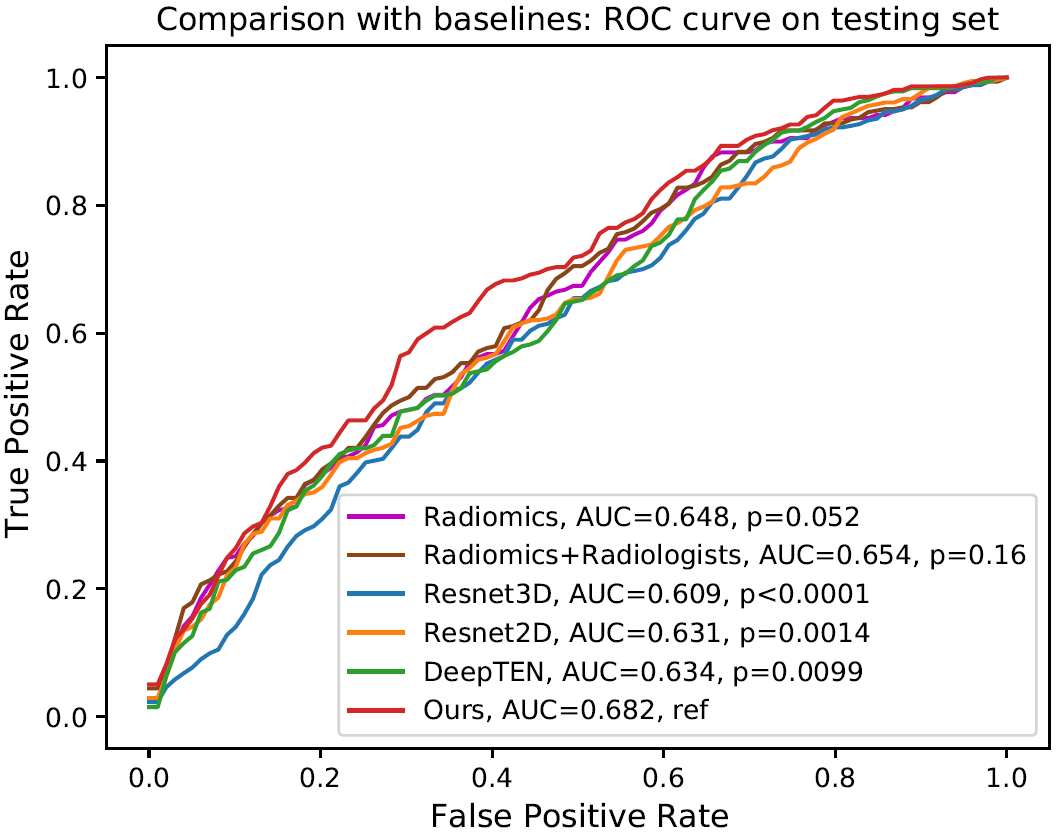}\label{Fig:BaselineROC}}
    \caption{ROC curve comparison of (\textbf{a}) ablation study and (\textbf{b}) baseline models and our method using nested five-fold cross-validation in Discovery dataset.}
    \label{Fig:ROC}
\end{figure*}

\textbf{Qualitative Evaluation.} 
Examples of LN segmentation and identification results are shown in Fig. \ref{Fig:LNSeg} (a) for qualitative comparison. Our segmentation model leverages prior knowledge of LNs' position distribution by incorporating the attention mechanism to remove false positives that are far from anatomically plausible LN areas. In Fig. \ref{Fig:LNSeg} (a), we can observe that nnUNet tends to falsely detect an instance inside some organs or located very far, while our method provides noticeably less false positives. 

\subsection{Evaluation of Patient-wise Lymph Node Metastasis Status Prediction} \label{Sec:metastasis}

\textbf{Metrics.} In this section, we evaluate various performance metrics of LN metastasis status prediction. For this binary classification problem, AUC, accuracy, balanced accuracy, sensitivity and specificity are adopted as evaluation metrics and the average results across 5 folds are reported. Statistical analysis is also carried out to verify the significance of performance improvement. 
We collect the predictions of all 5 folds, repeat 1,000 times of bootstrapping for calculating balanced accuracy, and apply Wilcoxon signed rank test to balanced accuracy distributoins to compare our method with several other configurations.
For comparing ROC curves, DeLong test is performed. $P$-values $\textless$ 0.05 are considered as statistically significant. To compute 95$\%$ CI, the 2.5th percentile and 97.5th percentile are estimated after 1,000 times of bootstrapping.

\textbf{Ablation Study.} We first investigate the impact of each component in our framework. To evaluate the metastasis prediction performance of LN segmentation and identification, the results can be aggregated into patient-level prediction, based on the definition that a patient with at least one positive LN is metastasis-positive. However, due to a large number of false positives produced by segmentation (LN segmentation in CT images is challenging after all), it will lead to a poor performance if predicting metastasis simply based on the presence of positive LN in the segmentation results. We instead conduct ROC analysis on the volume of the largest positive LN in each case, and find an optimal threshold with the best balanced accuracy in the validation set. Then the threshold is applied to the testing set. A patients with positive LNs larger than the threshold are classified into metastasis-positive; otherwise, it is classified into metastasis-negative. The ablation models  for consideration/comparison are listed as follows:

\begin{itemize}[leftmargin=*]
\setlength\itemsep{0em}
    \item ClsfromPDAC: The straightforward strategy combining ResNet2D \cite{he2016deep} feature and DeepTEN \cite{zhang2017deep} feature, extracted from PDAC slices, in the input of the classification layer.
    \item ClsbyLNSeg w/o Attn: Patient-level metastasis aggregation from the results of LN segmentation without attention (i.e. nnUNet).
    \item ClsbyLNSeg w/ Attn: Patient-level metastasis aggregation from the results of our proposed LN segmentation with attention.
    \item ClsfromPDAC + ClsbyLNSeg w/ Attn: Combined model incorporating the volume of the largest positive LN given by ClsbyLNSeg w/ Attn into the classification layer of ClsfromPDAC.
  
\end{itemize}

The results of the ablation experiments are summarized in Table \ref{Tab:LNMAbaltionBaseline}, and ROC analysis is illustrated in Fig. \ref{Fig:ROC}(a). By using only information about LNs, ClsbyLNSeg w/ Attn gives better aggregation results compared with ClsbyLNSeg w/o Attn (balanced accuracy 0.563 versus 0.545). Our final model (ClsfromPDAC + ClsbyLNSeg w/ Attn) significantly outperforms the other three models with a balanced accuracy of 0.633 ($p$-value $\textless$ 0.05), which reveals the success of integrating both tumor and LNs imaging information for metastasis prediction.

\textbf{Comparison with Baselines.} To validate the effectiveness of our method, radiomics model \cite{li2020contrast} and 2D/3D deep classification models are taken for comparison. To build the radiomics model, 1688 radiomics features of PDAC for each CT phase are extracted using Pyradiomics package \cite{van2017computational}\footnote{https://pyradiomics.readthedocs.io/}, and then shrunk using the least absolute shrinkage and selection operator (LASSO) method. Then a logistic regression model is applied to the selected features. The combined model of radiomics and CT-reported LN status is implemented with a logistic regression model on radiomics signature and radiologists' diagnosis. For 2D deep networks, ResNet2D \cite{he2016deep} and DeepTEN \cite{zhang2017deep}, we use ResNet-18 backbone pre-trained on ImageNet \cite{deng2009imagenet}; while for 3D deep networks, we adopt 3D-ResNet-18 \cite{hara3dcnns} backbone pre-trained on Kinetics-700 \cite{kay2017kinetics} and Moments in Time \cite{monfort2019moments}. In all of 2D/3D deep networks, a side branch with the PDAC mask as input is added to the backbone, as we implemented in our method, for fair comparison. Table \ref{Tab:LNMAbaltionBaseline} and Fig. \ref{Fig:ROC}(b) present the quantitative results of different models. More importantly, our method yields the best balanced accuracy (0.633) among all compared methods, and is significantly better than the radiomics method and all of 2D/3D deep networks. 

\subsection{External Validation}
In this section, we demonstrate the generalization ability of our LN metastasis status prediction in two external multi-center datasets (i.e., Ext-validation dataset 1 and Ext-validation dataset 2). After training the model on Discovery dataset using cross validation, we apply the model to external datasets for inference. For each patient, the ensemble prediction is generated by averaging the model predictions from five folds. We first evaluate the performance of ablation variants, and then compare our method with baseline models. Metrics are used the same as Section \ref{Sec:metastasis}.

\textbf{Ablation Study.}
We conduct ablation study on two external datasets, and results are shown in Table \ref{Tab:LNMAbaltionBaselineExt}. With respect to LN metastasis status prediction using only LN-related information, our method (ClsbyLNSeg w/ Attn) outperforms nnUNet (ClsbyLNSeg w/o Attn) on both two datasets (balanced accuracy 0.589 versus 0.579 on Ext-validation dataset 1, 0.639 versus 0.607 on Ext-validation dataset 2). By integrating PDAC characteristics, our final model (ClsfromPDAC + ClsbyLNSeg w/ Attn) gives the best results among all ablation models (balanced accuracy 0.620 on Ext-validation dataset 1 and 0.684 on Ext-validation dataset 2).

\begin{table*}[!ht]
\renewcommand\arraystretch{0.95}
\small
\centering
\caption{\small Performance comparison and ablation study on LN metastasis status prediction of two external datasets. Predictions are averaged across 5 folds. Wilcoxon signed rank test is conducted on balanced accuracy. * indicates $p$-value $\textless$ 0.05. NS indicates no significance.}
\label{Tab:LNMAbaltionBaselineExt}
\begin{tabular}{lllllll}

\toprule
\multirow{3}{*}{Dataset} & \multirow{3}{*}{Method} & Balanced &  \multirow{2}{*}{AUC} &  \multirow{2}{*}{Accuracy} &  \multirow{2}{*}{Sensitivity}&  \multirow{2}{*}{Specificity} \\
& & Accuracy & &  & &\\
&& [95$\%$ CI] & [95$\%$ CI] & [95$\%$ CI] & [95$\%$ CI] & [95$\%$ CI] \\
\midrule
 &\multirow{2}{*}{Radiomics } & 0.493$^*$ & 0.511 & 0.672 &  0.051 & \textbf{0.935}\\
&& [0.451-0.537] &[0.400-0.620] &  [0.626-0.710]&  [0.000-0.128] & [0.880-0.978] \\
&\multirow{2}{*}{ResNet3D} & 0.508$^*$& 0.509& 0.527&  0.462& 0.554\\
&& [0.415-0.612] &[0.409-0.617] & [0.450-0.611] &  [0.308-0.615] & [0.457-0.663] \\
&\multirow{2}{*}{ResNet2D} & 0.563$^*$&0.564& 0.542&  0.615& 0.511\\
&&[0.470-0.656]& [0.460-0.676] & [0.466-0.626] &   [0.462-0.769] & [0.413-0.609] \\
&\multirow{2}{*}{DeepTEN} & 0.556$^*$&0.557& 0.511&  \textbf{0.667}& 0.446 \\
{Ext-validation}&& [0.467-0.647] &[0.454-0.666] & [0.427-0.595] &  [0.513-0.795] & [0.348-0.544] \\
\cmidrule(lr){2-7}
{dataset 1} &\multirow{2}{*}{ClsfromPDAC} & 0.515$^*$& 0.554 & 0.485 &  0.590& 0.441\\
& & [0.423-0.609]& [0.450-0.661]& [0.402-0.568] & [0.436-0.744] & [0.344-0.548]\\
&\multirow{2}{*}{ClsbyLNSeg w/o Attn} & 0.579$^*$& 0.555 & 0.689&  0.308& 0.849\\
& & [0.498-0.662] & [0.454-0.641] & [0.621-0.750] &[0.179-0.462]&[0.774-0.914]\\
&\multirow{2}{*}{ClsbyLNSeg w/ Attn} & 0.589$^*$& 0.580& \textbf{0.705}&  0.308& 0.871\\
& &  [0.511-0.672]&[0.474-0.694]& [0.644-0.765]& [0.154-0.462]&[0.796-0.935]\\
&{Ours (ClsfromPDAC + } & \textbf{0.620}$^{ref}$ &\textbf{0.603}& 0.674& 0.487& 0.753\\
&{ClsbyLNSeg w/ Attn)}& [0.538-0.713] & [0.498-0.712] & [0.598-0.742] &  [0.333-0.641] & [0.667-0.839] \\

\midrule
 &\multirow{2}{*}{Radiomics } &0.508$^*$ & 0.609 & 0.441 &  0.243 & 0.773\\
&& [0.391-0.626] &[0.452-0.757] & [0.339-0.542] &  [0.108-0.378] & [0.591-0.909] \\
&\multirow{2}{*}{ResNet3D} & 0.461$^*$&0.442& 0.475&  0.514& 0.409\\
&&[0.334-0.584] & [0.300-0.593] & [0.356-0.594] &  [0.351-0.676] & [0.182-0.591] \\
&\multirow{2}{*}{ResNet2D} & 0.681$^*$&0.687& 0.650&  0.577& 0.786\\
&& [0.536-0.810] &[0.508-0.849] & [0.500-0.800] &  [0.385-0.731] & [0.571-0.930] \\
&\multirow{2}{*}{DeepTEN} & 0.613$^*$&0.647& 0.640&  \textbf{0.697}& 0.529 \\
{Ext-validation}&&[0.465-0.747] & [0.474-0.806] & [0.520-0.760] &  [0.545-0.848] & [0.294-0.765] \\
\cmidrule(lr){2-7}
 {dataset 2}&\multirow{2}{*}{ClsfromPDAC} & 0.620$^*$& 0.639 & 0.593&  0.514 & 0.727\\
& & [0.493-0.734] &[0.490-0.781] & [0.475-0.712] &[0.378-0.676]&[0.545-0.909]\\
&\multirow{2}{*}{ClsbyLNSeg w/o Attn} & 0.607$^*$&0.690& 0.542&  0.351& \textbf{0.864}\\
&&[0.503-0.716]&[0.554-0.818] & [0.441-0.661] &  [0.216-0.487]&[0.682-1.000]\\
&\multirow{2}{*}{ClsbyLNSeg w/ Attn} & 0.639$^*$& 0.695 & 0.593&  0.459& 0.818\\
& &[0.525-0.752]& [0.552-0.833] & [0.475-0.695]& [0.297-0.622]&[0.636-0.955]\\
&{Ours (ClsfromPDAC + } &\textbf{0.684}$^{ref}$ & \textbf{0.703} & \textbf{0.661} & 0.595& 0.773\\
&{ClsbyLNSeg w/ Attn)}& [0.570-0.797] &[0.554-0.846] & [0.542-0.780] &  [0.432-0.757] & [0.591-0.909] \\
\bottomrule
\end{tabular}
\vspace{-1em}
\end{table*}

\textbf{Comparison with Baselines.} Table \ref{Tab:LNMAbaltionBaselineExt} validates the generalization performance of our method compared with radiomics and 2D/3D deep learning models. Note that we skip methods involved with CT-reported LN status since there is no CT report available in two external datasets. The radiomics model shows poor generalization ability with a large drop in performance compared with that in Table \ref{Tab:LNMAbaltionBaseline}, while deep learning models are relatively more robust. Our method significantly surpasses all of 2D/2D deep learning models with $p$-value $\textless$ 0.05 on both external datasets (balanced accuracy 0.620 and 0.684 respectively), demonstrating the power of our model to generalize across different data sites.  

\section{Discussion}\label{sec:discussion}

Pre-operative LN metastasis status prediction is of vital significance for PDAC patients in the following aspects. Firstly, if diagnosed with LN metastasis, patients with resectable PDAC are recommended to receive neoadjuvant therapy first before surgery, according to NCCN guidelines \cite{tempero2021pancreatic}. Secondly, pancreatectomy could be guided by whether and where their LNs have metastasized, that is, whether or not a standard or an extended lymphadenectomy should be performed. This could make the surgical procedure being more targeted beforehand which could lead to better patient outcome and avoid over-treatment. Thirdly, LN metastasis is highly associated with patients' survival, which can evidently assist with good prognosis prediction value \cite{yao2021deepprognosis}. Note that it is very time consuming and highly dependent on (board-certified radiologist) observer's experience and energy level to manually determine whether a patients has LN metastasis primarily from CT scans (even it is a very desirable task for patient care). CT-reported LN status in this study shows limited performance with an accuracy of 0.599, thus accurate LN metastasis status prediction is highly desired.

In the literature, LN metastasis status prediction has predominantly been studied through tumor feature extraction, combined with CT report information, using radiomics  ~\cite{ji2019biliary,wang2020ct,li2020contrast,bian2019relationship,yang2020integrating,gao2020radiomics,meng20202d} or deep learning approaches~\cite{jin2021deep, dong2020deep}, while the one leveraging LN radiomics requires manual delineation and considers simply the LN with the largest size~\cite{yang2020integrating}. An automated and accurate
process of LN segmentation and nodal positivity identification is hence of high importance for assisting metastasis prediction. Predicting the metastasis status from automated segmented LNs is formulated by detecting metastatic LNs with Faster R-CNN \cite{lu2018identification}, however the spatial context priors towards LNs are not exploited. This work proposes an automated geometric attention mechanism using LN segmentation and identification to predict the patient-level status of LN metastasis. 

To demonstrate the effectiveness of our method, we provide extensive quantitative experiments on LN segmentation/identification and LN metastasis status prediction. Our LN segmentation model statistically significantly outperforms the strong baseline nnUNet in voxel-wise and instance-wise metrics. For LN instance-wise detection, our model achieves considerable quantitative improvements (4.6$\%$) in precision (with respect to a similar recall level) as compared to nnUNet (see Table \ref{Tab:LNSeg_Results}). This observation clearly validates that the proposed distance-guided attention mechanism is beneficial to remove false positives as we expect. The success of our model can be attributed to its attention map design and informative negative selection scheme. The former defines the LN-plausible regions that deserve network's focus, and the latter helps to throw out non-informative negative training patches accordingly. As such, it becomes more efficient to train and force the model to learn discriminative features from possible LN locations. To verify the effect of LN detection improvements on patient-level metastasis status prediction, we perform instance-wise positivity identification and patient-wise aggregation on the LN instances to classify the patients into metastasis-positive/-negative, and our model presents better prediction performance than nnUNet (balanced accuracy 0.563 versus 0.545, Table \ref{Tab:LNMAbaltionBaseline}). We further combine the results with tumor CT imaging characteristics and our final prediction model achieves statistically significant performance gains compared to radiomics methods and other deep 2D/3D models (see Table \ref{Tab:LNMAbaltionBaseline}), which demonstrates the success and effectiveness of integrating tumor morphology and lymphatic anatomy. It is worth noting that our method achieves statistically significant improvement (balanced accuracy 0.633 versus 0.604) compared to the approach even with radiologists involved in ``Radiomics + CT-reported LN status" in Table \ref{Tab:LNMAbaltionBaseline}. Nevertheless, using our method, this time-consuming, subjective and highly challenging manual process of CT-reported LN status can be fully automated. External multi-center clinical validation is further conducted on extra two datasets from different hospitals, and the results evidently exhibit our superior performance accuracy and generalization ability with the best results (balanced accuracy 0.620 and 0.684 on the two external datasets) among several compared models (see Table \ref{Tab:LNMAbaltionBaselineExt}). With all above-mentioned experiments, our model reports highly generalized prediction performance (0.620$\sim$0.684) on multi-center datasets and robust improvements over CT-reported LN status (0.599) as well as radiomics and deep learning models, which clearly clarifies the advantage and stability of our model.

Although recent work \cite{li2020contrast,gao2020radiomics} report exceedingly high accuracy (AUC $\textgreater$ 0.9), they use small datasets of $<200$ patients, which would be subject to overfitting. Another recent progress in gastric cancer \cite{meng20202d} enrolls over 500 patients from multiple hospitals, and yields noticeably lower but probably more reliable AUC score of 0.615$\sim$0.712 in validation using 2D/2.5D/3D radiomics features and under different patient splits. \cite{meng20202d} is probably more suitable to serve as reference baseline for our work. We employs 940 patients in total in this study, in which 749 patients are from a high-volume pancreatic cancer clinical center and 191 are from two external centers. The studied patient population is arguably much closer and more realistic to the real-world patient data distributions comparing to \cite{li2020contrast,gao2020radiomics}, similar to \cite{meng20202d}. We present a very promising approach that explicitly explores the role of automated LN segmentation in promoting LN metastasis status prediction to facilitate future clinical adoption as a fully-automated and generalizable clinical tool. One limitation of our framework lies in the intuitive but simple solution that extracts tumor and LNs imaging information separately and then integrates them by feature concatenation, which does not fully exploit the nature of interactions between tumor and cancer cells in LNs. This work could be further improved in the future by designing an enriched deep learning geometric network representation to encode the tumor-LN topology information and spatial anatomical interactions, by modeling the pathways of nodal metastasis explicitly. 

Last, without loss of generality, our proposed method is applicable for finding the preoperative LN metastasis status of other types of solid tumor or cancers, such as liver or gastric cancers. We leave this as future work.

\section{Conclusion}
We present an attention based LN segmentation network and utilize it on predicting LN metastasis in patients with PDAC. The proposed LN segmentation network involves an attention mechanism that encourages the network to focus on regions around certain anatomical organs/vessels. It outperforms the strong baseline nnUNet\cite{isensee2021nnu} by leveraging the context information of surrounding anatomical structures. Our segmentation model, followed by a nodal positivity identification model, can serve as a single predictor for LN metastasis. Combined with tumor imaging characteristics, we further build a compositive LN metastasis status prediction model that is validated to surpass the CT-reported results, radiomics based method, and other 2D/3D deep learning models. Further investigations include conceiving a more complicated way to encode tumor-LN relationship and exploring its applications to prognosis and treatment planning in cancer patient management.

\section*{Acknowledgments}
This work was supported in part by the National
Science Foundation for Scientists of China (81871352,
82171915, and 82171930), Clinical Research Plan of SHDC
(SHDC2020CR4073), 234 Platform Discipline Consolidation
Foundation Project (2019YPT001, 2020YPT001), and The
Natural Science Foundation of Shanghai Science and Technology Innovation Action Plan (21ZR1478500, 21Y11910300).

\input{refs.bbl}

\bibliographystyle{IEEEtran}
\bibliography{refs}

\newpage

 





\end{document}